\documentclass[nonacm,sigconf]{acmart}
\renewcommand\footnotetextcopyrightpermission[1]{\ignorespaces} 
\usepackage{diagbox}
\usepackage{subfiles}
\usepackage{style/customize}
\usepackage{physics}
\usepackage{threeparttable}
\usepackage{multirow}

\AtBeginDocument{%
  \providecommand\BibTeX{{%
    \normalfont B\kern-0.5em{\scshape i\kern-0.25em b}\kern-0.8em\TeX}}}

\setcopyright{none}
\begin{document}

\title[]{SFQ counter-based precomputation for large-scale cryogenic VQE machines}

\author{Yosuke Ueno}
\affiliation{%
  \institution{RIKEN}
  \state{Saitama}
  \country{Japan}}
\email{yosuke.ueno@riken.jp}

\author{Satoshi Imamura}
\affiliation{%
  \institution{Fujitsu Limited}
  \state{Kanagawa}
  \country{Japan}}
\email{s-imamura@fujitsu.com}

\author{Yuna Tomida}
\affiliation{%
 \institution{The University of Tokyo}
 \state{Tokyo}
 \country{Japan}}
\email{tomida@hal.ipc.i.u-tokyo.ac.jp}

\author{Teruo Tanimoto}
\affiliation{%
 \institution{Kyushu University}
 \state{Fukuoka}
 \country{Japan}}
\email{tteruo@kyudai.jp}

\author{Masamitsu Tanaka}
\affiliation{%
 \institution{Nagoya University}
 \state{Aichi}
 \country{Japan}}
\email{masami\_t@ieee.org}

\author{Yutaka Tabuchi}
\affiliation{%
  \institution{RIKEN}
  \state{Saitama}
  \country{Japan}}
\email{yutaka.tabuchi@riken.jp}

\author{Koji Inoue}
\affiliation{%
 \institution{Kyushu University}
 \state{Fukuoka}
 \country{Japan}}
\email{inoue@ait.kyushu-u.ac.jp}

\author{Hiroshi Nakamura}
\affiliation{%
 \institution{The University of Tokyo}
 \state{Tokyo}
 \country{Japan}}
\email{nakamura@hal.ipc.i.u-tokyo.ac.jp}

\renewcommand{\shortauthors}{}

\begin{abstract}

The variational quantum eigensolver (VQE) is a promising candidate that brings practical benefits from quantum computing. However, the required bandwidth in/out of a cryostat is a limiting factor to scale cryogenic quantum computers.
We propose a tailored counter-based module with single flux quantum circuits in 4-K stage which precomputes a part of VQE calculation and reduces the amount of inter-temperature communication.
The evaluation shows that our system reduces the required bandwidth by 97\%, and with this drastic reduction, total power consumption is reduced by 93\% in the case where 277 VQE programs are executed in parallel on a 10000-qubit machine.

\end{abstract}



\maketitle

\section{Introduction\label{sec:introduction}}

Quantum computers (QCs) require classical computers (CCs) and frequent interactions with them not only for control and management of quantum processing units (QPUs) but also for higher-level processing.
Examples of such interactions include quantum error correction in fault-tolerant quantum computers~(FTQC) and classical parameter optimization of variational quantum algorithms~(VQAs) in noisy intermediate-scale quantum (NISQ) computers.
Consequently, QCs need to ensure sufficient bandwidth between a QPU and CCs for essential interactions.

\begin{figure}[t]
  \centering
  \includegraphics[width=\columnwidth]{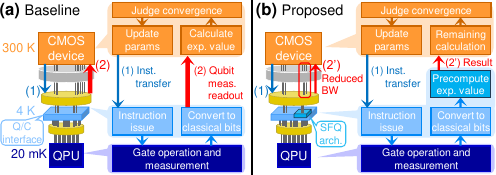}
  \vspace{-4mm}
  \caption{(a) Baseline and (b) proposed system designs of SQC with their VQE procedures.}
  \label{fig:cryo-vqe}
\end{figure}

Superconducting QCs~(SQCs) are leading the QC development, where the qubits operate in a cryostat, as shown in Fig.~\ref{fig:cryo-vqe}, because of their susceptibility to thermal noise. 
SQCs use high-frequency coaxial cables for inter-temperature communication between a room-temperature CC and a cryogenic QPU. 
To keep QPU work correctly, the power dissipation~(\textit{i.e.}, power consumption of QPU and peripherals, and heat inflow through cables connecting temperature stages) inside the cryostat has to be within its cooling capacity~\cite{ueno2023inter}.
The amount of heat inflow depends on the required bandwidth between in and out of the cryostat, with parameters of the material, the diameter, and the number of cables.
Currently, the number of qubits is increasing to execute larger-scale programs and to explore an FTQC system.
It will be more important to carefully design the interaction to suppress the power dissipation inside the cryostat.

Variational quantum eigensolver~(VQE) gathers a significant interest as a usecase of QCs because it can be used for quantum chemistry problems, which are known to be difficult to solve with CCs.
The expectation value calculation is a main process of VQE and can be performed by counting qubit measurement results according to a target Hamiltonian.
As shown in Fig.~\ref{fig:cryo-vqe}~(b), we propose to perform a part of the expectation value calculation of VQE in advance using single-flux quantum~(SFQ) circuits located in the 4 K stage of the cryostat, aiming to reduce the required bandwidth between in/out of the cryostat.
In our previous work, we have proposed a similar architecture targeting quantum approximate optimization algorithm~(QAOA)~\cite{ueno2023inter}; however, the architecture cannot be applied for VQE because it cannot calculate the expectation values of arbitrary Hamiltonian terms.
Our contributions are summarized as follows.

\begin{enumerate}
\item We analyze the inter-temperature communication of VQE to identify the bandwidth bottleneck~(Sec.~\ref{analysis}).
\item We propose the precomputation architecture to reduce the inter-temperature bandwidth of cryogenic VQE machines and design it with SFQ circuits (Sec.~\ref{proposal})
\item We discuss the trade-off between the power dissipation of cables and that of our precomputation architecture on single- and multi-program situations (Sec.~\ref{evaluation:scalability}).
\end{enumerate}

\section{Background}
\label{background}

\subsection{Variational Quantum Eigensolver~(VQE)}
\label{background:vqe}

VQE is a variational hybrid quantum-classical eigensolver designed for NISQ machines and can be used to calculate the ground-state energy of a target molecule in quantum chemical calculations~\cite{perruzo2014vqe}. 
It calculates an approximate ground-state energy by iteratively executing a parameterized quantum circuit on a quantum device and updating its parameters using a classical optimizer until the expectation value calculated from the measurement result of the quantum circuit converges to a threshold. 
We summarize the notations used throughout this paper in Tab.~\ref{tab:notation}.

\subsubsection{Expectation value calculation in VQE}
\label{background:vqe-exp}
VQE obtains ground-state energy by finding the optimal parameter $\boldsymbol{\theta}$ whose generated quantum state $\ket{\psi{(\boldsymbol{\theta})}}$ minimizes the expectation value of a Hamiltonian $H$, $\expectation{H}$. 
A Hamiltonian is encoded to a linear combination of Pauli terms through second quantization and a transformation method such as Jordan-Wigner~\cite{Fradkin:1989jo}, as shown in $H = \sum_{i}w_{i}P_{i}$, where $w_{i}$ is a weight and $P_{i}$ is a Pauli string. 
The expectation value of $P_{i}$, $\expectation{P_{i}}$, is calculated from $T$ times measurement results of a parameterized quantum circuit constructed based on an \emph{ansatz}. 
We call this $T$-times measurement procedure \emph{Pauli loop} in this paper.
Since each qubit can be directly measured on the $Z$-basis, $X$- or $Y$-basis measurement is realized by basis transformation with a Hadamard or RX($\pi/2$) gate before $Z$-basis measurement, respectively.  
For instance, to calculate $\expectation{X_0Z_1Z_2Y_3}$, we prepare $\ket{\psi'(\boldsymbol{\theta})} = H_0RX(\pi/2)_3\ket{\psi(\boldsymbol{\theta})}$ and measure it on the $Z$-basis.
Based on the $Z$-basis measurement, $\expectation{P_{i}}$ can be calculated as 
\begin{align}
\expectation{P_{i}} = (C_{i, even}-C_{i, odd})/T = (T-2C_{i, odd})/T, \label{eq:pauli_string_exp}
\end{align}
where $C_{i, odd}$ ($C_{i, even}$) represents the total count of measurement results where the sum of the values of target qubits is odd (even) among one Pauli loop. 

\subsubsection{Group measurement (GM)}
\label{background:vqe-gm}
Naively, $\expectation{H}$ can be calculated by measuring a parameterized quantum circuit per Pauli string $P_{i}$ and summing up all $\expectation{P_i}$, each multiplied by the corresponding weight $w_{i}$. 
However, this is time-consuming because the quantum circuit must be executed the same number of times as the number of Pauli strings $N_P$, which grows as $O(N^4)$~\cite{Verteletskyi:2020me}. 
GM is a well-known technique to assign Pauli strings to a group $G$ as many as possible and realize multiple Pauli string measurements in a single quantum circuit execution. 
In this paper, $GP$ denotes an inclusive Pauli string covering all of Pauli strings in a group $G$.
For instance, for $G = \{I_0X_1Z_2, X_0X_1I_2, I_0I_1Z_2\}$, $GP$ is $X_0X_1Z_2$.
In addition, we denote the number of groups by $N_G$ and the maximum number of Pauli strings in a group, which is called \emph{group size}, by $K$.

Recent grouping methods significantly reduced the number of quantum circuit executions by efficiently grouping Pauli strings, which means grouping results with smaller $N_G$~\cite{Verteletskyi:2020me, Kurita:2023pa}. 
Note that these methods can reduce $N_G$ to $\order{N^4}$~\cite{Verteletskyi:2020me} or $\order{N}$~\cite{Kurita:2023pa}. 

\begin{table}[tb]
\caption{Notations about VQE used in this paper}
\label{tab:notation}
\scriptsize
\begin{tabular}{|c|l||c|l|} \hline
Notation                   & Description                      & Notation                   & Description                                                  \\ \hline
$N$                        & \# qubits of machine    &  $\ket{\psi}$              & Ansatz state                                                 \\
$T$                        & \# shots                         &  $\boldsymbol{\theta}$     & Ansatz parameters~\cite{kandala2017hardware}                  \\
$H$                        & Hamiltonian                      &  $D$                       & Depth of hardware efficient anzats~\cite{kandala2017hardware} \\
$P$                        & Pauli string                     &  $G$                       & Group of Pauli strings                                       \\ 
$N_P$                      & \# Pauli strings                 &  $GP_i$                    & Grouping Pauli string for group $G_i$                        \\ 
$w$                        & Weight of a Pauli string         &  $N_G$                     & \# Groups                                                    \\ 
                           &                                  &  $K$                       & Group size (Max \# Pauli strings in a group)                 \\ \hline
\end{tabular} 
\begin{tablenotes}
  \item[*] \# in the table means "the number of."
\end{tablenotes}
\end{table}

\subsection{Single flux quantum (SFQ) logic}
\label{background:sfq}
SFQ logic is a digital circuit that uses superconductor devices.
At the core of its functionality, the SFQ processes data through magnetic flux quanta, which are held within superconductor loops. 
These loops incorporate Josephson junctions (JJs), serving pivotal roles as the switch mechanisms.
The presence and absence of a single magnetic flux quantum 
in the loop represent logical `1' and `0', respectively. 

Recently, SFQ logic has been attractive as peripheral circuits for cryogenic QCs because of its high-speed operation, power efficiency, and ability to operate in a cryogenic environment~\cite{ueno2021qecool,jokar2022digiq}.
In this paper, we use SFQ logic to design our cryogenic architecture for optimized inter-temperature communication of SQCs.

\subsection{Related work on inter-temperature bandwidth reduction in SQCs}
Jokar \textit{et al.}~\cite{jokar2022digiq} proposed an SFQ-based qubit control system based on the SIMD concept. 
While their system focuses on the qubit controlling communication from room- to low-temperature environment, ours concentrates on the measurement readouts by precomputation of the expectation value of VQE in a cryogenic environment.
In the ﬁeld of physics, Opremcak \textit{et al.}~\cite{opremcak2021measurement} proposed a method for efficient qubit measurements in cryogenic environments by equipping photon detectors on an SFQ chip. 
Based on such a fundamental technique, we present a system architecture for reducing inter-temperature communications for qubit measurements.

Previously, we proposed an SFQ-based architecture for inter-temperature bandwidth reduction in SQCs~\cite{ueno2023inter}, and this study builds upon the previous work. 
In the previous paper, we focused on QAOA, which is a simple case of VQE, and achieved bandwidth reduction with a simple SFQ architecture that merely \emph{counts} qubit measurement values in a cryogenic environment.
However, our previous architecture is not applicable to other VQAs, including VQE, due to its simplified design for QAOA.
In this paper, we make a breakthrough by removing the critical limitation of Ref.~\cite{ueno2023inter} by introducing an SFQ-based precomputation technique.

\section{Inter-temperature bandwidth modeling of VQE machines\label{analysis}}
\subsection{Baseline system design}
Figure~\ref{fig:cryo-vqe}~(a) shows the baseline system. 
We assume that the system has a quantum-classical interface at the 4-K stage; an SFQ interface generates SFQ pulses to control and measure qubits in QPU and detects `0' or `1' depending on the state of each qubit. 
The execution flow of VQE in the baseline system is a repetition of the cycle shown on the right side of Fig.~\ref{fig:cryo-vqe}~(a). 
Inter-temperature communications occur during (1) the instruction transfer to QPU control and (2) the readout of qubit measurement results, as shown in Fig.~\ref{fig:comm-timing}.
In the following subsections, we model the required bandwidths of these communications on the baseline system and discuss their major bottleneck.
Here, the term ``required bandwidth'' in this paper refers to sufficient bandwidth from/to the cryogenic environment to prevent communication delays during the execution of quantum circuits.

\subsection{Bandwidth for inter-temperature communication}
\label{analysis:bandwidth}
\begin{figure}[t]
  \centering
  \includegraphics[width=\columnwidth]{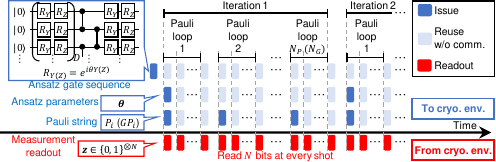}
  \vspace{-4mm}
  \caption{Inter-temperature communication timing in baseline VQE machine without (with) GM}
  \label{fig:comm-timing}
\end{figure}

In VQE applications, the execution time of the quantum circuit $t_{QC}$ highly depends on the chosen ansatz. 
In this paper, we assume a hardware efficient ansatz~\cite{kandala2017hardware}.
The quantum circuit of VQE repeats the same sequence of gates differing in Pauli string $P_i$ for basis transformation and ansatz parameters $\boldsymbol{\theta}$ following the given Hamiltonian. 
The left top area of Fig.~\ref{fig:comm-timing} shows the gate sequence of the hardware efficient ansatz.

\subsubsection{Case without GM\label{subsubsec:bandwidth_without_GM}}
First, we model the required bandwidth of the baseline for (1) instruction transfer and (2) measurement readout without GM.

\noindent
\textbf{(1) Bandwidth for instruction transfer: }
As instruction transfer communication, we need to send three types of information to the cryogenic environment: ansatz gate sequence, ansatz parameters, and Pauli string.
In VQE computation, each iteration has the same ansatz gate sequence differing only in ansatz parameters $\boldsymbol{\theta}$. 
Therefore, it is transferred only once before the first iteration and reused through the VQE computation, as shown in Fig.~\ref{fig:comm-timing}.
By transferring the gate sequence at low speed, we can ignore its impact on the inter-temperature bandwidth with negligible execution time overhead.

The ansatz parameters $\boldsymbol{\theta} = \lbrace \theta_k^{q, d}\rbrace_{k=1, 2}^{q=1, ..., N, d=1, ..., D+1}$ are transferred once per circuit parameter update at every iteration, \textit{i.e.}, once every $N_pT$ shots, where $N_p$ is the number of Pauli strings. 
Suppose each parameter has $b_\theta$ bitwidth, the required bandwidth for transferring ansatz parameters $\BWansatzparams$ is modeled as
\begin{align}
    \BWansatzparams = \frac{2N(D+1)b_\theta}{N_PTt_{QC}} \sim \order{1/N^3T} \hspace{3mm} \left(\because N_P\sim\order{N^4}\right). \label{eq:bw_ap}
\end{align}

By contrast, we need to transfer a Pauli string once for each Pauli loop.
Since one Pauli string is a $2N$-bit information, which indicates $I/X/Y/Z$ for each of $N$ qubits,
the required bandwidth for Pauli string $\BWpaulistring$ is modeled as
\begin{align}
    \BWpaulistring = 2N/Tt_{QC} \sim\order{N/T}.  \label{eq:bw_ps}
\end{align}

Based on the discussion above, Eqs.~(\ref{eq:bw_ap}) and (\ref{eq:bw_ps}) show that $\BWpaulistring$ is dominant in the instruction transfer bandwidth.
Then, we model the required bandwidth for the instruction transfer $\BWInst = \BWpaulistring$.

\noindent
\textbf{(2) Bandwidth for measurement readout: }
The required bandwidth for read-out $\BWMeas$ can be estimated as $ N $ bits per shot, as shown in Fig.~\ref{fig:comm-timing},
\textit{i.e.},
\begin{align}
  \BWMeas = N / t_{QC}. \label{eq:bw_measure} 
\end{align}

\subsubsection{Case with GM\label{subsubsec:bandwidth_with_GM}}
Next, we model the required bandwidth of the baseline system for the case with GM and show the same discussion with Sec.~\ref{subsubsec:bandwidth_without_GM} applies to this case.

\noindent
\textbf{(1) Bandwidth for instruction transfer: }
In the baseline system with GM, while we perform computations for up to $K$ Pauli strings simultaneously, the required bandwidth for instruction transfer is the same as in the case without GM for the reasons given below.
While the ansatz parameters are the same in the case without GM, the communication interval changes because the number of shots in one iteration changes from $TN_P$ to $TN_G$ by GM.
Thus, $\BWansatzparams$ is obtained by replacing $N_P$ in Eq.~(\ref{eq:bw_ap}) with $N_G$. 
Regarding Pauli strings, when performing GM in the baseline system, an inclusive Pauli string $GP$ for a group $G$ is transferred, instead of sending all Pauli strings in $G$.
Since $GP$ is a $2N$-bit information as same as $P$, the required bandwidth for Pauli string $\BWpaulistring$ is the same as Eq.~(\ref{eq:bw_ps}).
As a result, since the order of $N_G$ is $\order{N^4}$~\cite{Verteletskyi:2020me} or $\order{N}$~\cite{Kurita:2023pa}, $\BWpaulistring$ dominates in the instruction transfer bandwidth, similar to the case without GM.

\noindent
\textbf{(2) Bandwidth for measurement readout: }
In the baseline system, the procedure of transmitting $N$-bit qubit measurements for each shot is the same in both cases without and with GM. 
Therefore, $\BWMeas$ is the same as expressed in Eq.~(\ref{eq:bw_measure}).

\subsection{Analysis on bandwidth bottleneck}
\label{analysis:design-time-bandwidth}
Based on the bandwidth models in Sec.~\ref{analysis:bandwidth}, we identify the inter-temperature bandwidth bottleneck of the VQE machine.
In the baseline system, the ratio of $\BWInst$ to $\BWMeas$ is calculated as
\begin{align}
    \frac{\BWInst}{\BWMeas} = \frac{2N}{Tt_{QC}} \cdot \frac{t_{QC}}{N} 
    = 2/T \ll 1. 
    \hspace{3mm}
    (\because T\sim10^6) \label{eq:bw_bottleneck}
\end{align}
Here, $T\sim10^6$ is used as the number of shots required to achieve the ``chemical accuracy''~\cite{perruzo2014vqe}. 
Therefore, measurement readout is the dominant factor in the inter-temperature bandwidth of the baseline system.

While the discussion above is based on a single-program situation, note that in multi-program scenarios, Eqs.~(\ref{eq:bw_ap})---(\ref{eq:bw_measure}) simply get multiplied by the program parallelism factor $L$\footnote{
In a specific case where the $L$ programs share a common set of Pauli strings, Eq.~(\ref{eq:bw_measure}) is multiplied by $L$ while Eq.~(\ref{eq:bw_ps}) remains the same, hence the impact of measurement readout on the required bandwidth further increases.}.
Hence the same conclusion is derived from Eq.~(\ref{eq:bw_bottleneck}) in multi-program scenarios.

\section{Precomputation at the 4-K stage}
\label{proposal}

\subsection{Precomputation without GM\label{subsec:precomputation_circuit}}
As shown in Fig.~\ref{fig:cryo-vqe}~(a), the baseline system calculates $\expectation{H}$ in a 300-K environment based on the qubit measurements. 
Thus, we need inter-temperature communications to send measurement results of $N$ qubits after each shot ends, and the required bandwidth of the baseline system is $\BWMeas = N/t_{QC}$ as explained in Sec.~\ref{analysis}.

By contrast, our system computes a part of the $\expectation{H}$ in a 4-K environment to reduce the bandwidth, as shown in Fig.~\ref{fig:cryo-vqe}~(b).
Our architecture calculates $C_{i, odd}$ in Eq.~(\ref{eq:pauli_string_exp}) for a given Pauli string $P_i$ as follows.
Let us denote $PM_i\in \{0, 1\}^N$ as a Paulimask, which indicates non-$I$ positions of given $P_i$, and $\mathbf{z}^t \in\{0, 1\}^N$ as $t$-th measurement result of $N$ qubits, respectively.
Our architecture calculates $c_{i, t} \in \{0, 1\}$ as 
\begin{align}
    c_{i, t} = PM_i \cdot \mathbf{z}^t \pmod2
\end{align}
in a 4-K environment, and $C_{i, odd}$ is calculated with $c_{i, t}$ as $C_{i, odd} = \sum_t c_{i, t}$ in a room-temperature environment.
For instance, for Pauli string $P_i = X_0I_1Z_2Y_3$ and $\mathbf{z}^t = \{1, 0, 1, 0\}$, $c_{i, t}$ is calculated as $c_{i, t} = 1\cdot1 \oplus 0\cdot0 \oplus 1\cdot1 \oplus 1\cdot0 = 0$.
Therefore, by sending $c_{i, t}$ instead of $\mathbf{z}^t$ for each shot, the required bandwidth $\BWProposed = \BWMeas/N = 1/t_{QC}$ becomes a constant to $N$.
Note that no execution time overhead is introduced by our method for the case without GM.

\subsection{Counter-based precomputation with GM\label{subsec:couter_with_GM}}
When performing GM in the VQE procedure, we simultaneously compute the expectation values for up to $K$ Pauli strings.
Therefore, if we naively apply the proposed method, the required bandwidth of our system $\BWProposed$ becomes $K/t_{QC}$, and it may exceed that of the baseline system depending on the value of $K$.
Here, we apply the counter-based approach of the previous work for QAOA~\cite{ueno2023inter} to reduce the required bandwidth for the case with GM.

\subsubsection{MSB-sending policy}
We assume that our architecture has the $K$ precomputation circuits mentioned in Sec.~\ref{subsec:precomputation_circuit} and $K$ $b$-bit counters for group size $K$.
As in the previous study~\cite{ueno2023inter}, $c_{i, t}$ values for all Pauli strings in a given group are counted with counters in a 4-K environment through a Pauli loop.
Only the MSBs of counters are transferred to the room-temperature environment each time a counter overflows, thereby reducing the required inter-temperature bandwidth by a factor of $1/2^{b-1}$ to that of the baseline system.
As a result, in our proposed system, the bandwidth requirement for MSB is $\BWMSB = K/(2^{b-1}t_{QC})$ for the case with GM.

\subsubsection{Collecting non-MSBs\label{subsec:non-MSB}}
\begin{figure}[tb]
  \centering
  \includegraphics[width=\columnwidth]{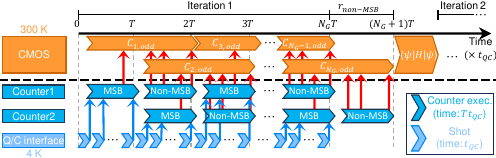}
  \vspace{-4mm}
  \caption{Inter-temperature communication in VQE using proposed architecture}
  \label{fig:MSB_nonMSB_send_timing}
\end{figure}
In our method, while the MSB of each counter is sent during a Pauli loop, we need to collect $Kb$ non-MSBs of the counters to a room-temperature environment after the Pauli loop for the calculation of expected values, as in the previous work for QAOA~\cite{ueno2023inter}. 
However, in contrast to the QAOA case, where $(i+1)$-th iteration cannot start until the non-MSB collecting for $i$-th iteration, in VQE, we can start for $G_{i+1}$-th Pauli loop before the non-MSB collecting for $G_i$.
Hence, by performing the non-MSB collection for $G_i$ and $G_{i+1}$-th Pauli loop simultaneously, as shown in Fig.~\ref{fig:MSB_nonMSB_send_timing}, most of the latency caused by the non-MSB collection can be hidden, resulting in its minor impact to the total execution time.

To hide the non-MSB latency, we prepare two counters for one Pauli string.
Each counter is used alternately for MSB sending and non-MSB collection, as shown in Fig.~\ref{fig:MSB_nonMSB_send_timing}.

Based on the discussion above, we model the required bandwidth $\BWnonMSB$ and execution time overhead to the total execution time $r_{\text{non-MSB}}$.
Suppose that we collect the non-MSBs within $Tt_{QC}$ as shown in Fig.~\ref{fig:MSB_nonMSB_send_timing}, $\BWnonMSB$ equals to $Kb/(Tt_{QC})$.
To keep $\BWnonMSB$ as small as $\BWMSB$, b must satisfy 
\begin{align}
 \BWnonMSB \leq \BWMSB 
  \hspace{2mm} \Leftrightarrow \hspace{2mm}
b 2^{b-1} \leq T.     
\end{align}

As shown in Fig.~\ref{fig:MSB_nonMSB_send_timing}, the total execution time of each iteration with and without our method is $N_GTt_{QC}$ and $(N_G+1)Tt_{QC}$, respectively.
Hence, $r_{\text{non-MSB}}$ is calculated as 
\begin{align}
1 + r_{\text{non-MSB}} = 
\frac{(N_G+1)Tt_{QC}}{N_GTt_{QC}}
  \hspace{2mm} \Leftrightarrow \hspace{2mm}
 r_{\text{non-MSB}} = 1/N_G.
\end{align}
Note that the order of $N_G$ to $N$ is $\order{N^2}$, as explained in Sec.~\ref{background:vqe-gm}.
As a result, the execution time overhead of our architecture is negligible as $N$ scales.

\subsubsection{Instruction transfer overhead}
To compute $c_{i, t}$ values for all Pauli strings in a group $G$ simultaneously, our system requires up to $K$ separate Paulimasks; however, these cannot be obtained from an inclusive Pauli string $GP$.
Therefore, we need to transfer $K$ Paulimasks for our system instead of one $GP$, which increases instruction transfer bandwidth compared to the baseline.
Since each Paulimask is an $N$-bit information, the Pauli string transfer bandwidth for the proposed system with GM $\BWpaulistring'$ is modeled as 
\begin{align}
  \BWpaulistring' = \frac{KN}{Tt_{QC}}.  \label{eq:bw_proposed_ps}
\end{align}
To discuss the required bandwidth reduction, we need to put an additional assumption to $K$ and $T$ with consideration of a practical environmental setup, \textit{i.e.}, $N < 10^4$, $K\sim \order{N^2}$, and $T \sim 10^6$.
Therefore, the condition where the increased bandwidth for instruction transfer is not dominant in the proposed system with GM is 
\begin{align}
    \BWpaulistring'/\BWMeas \ll 1
    \Leftrightarrow
    K \ll T.
    \label{eq:KT_condition}
\end{align}
The remainder of this paper assumes Eq.~(\ref{eq:KT_condition}) unless explicitly noted.

\subsection{Architecture design with SFQ circuit}
\label{proposal:counter}

\begin{table}[tb]
\caption{Detailed configuration of the proposed circuit based on the AIST 10-kA/$\text{cm}^2$ ADP cell library\cite{detail_of_cell_library_ADP2}}
\label{tab:circuit_configuration}
\scriptsize
\begin{tabular}{|c|c|c||c|c|c|c|c|c|} \hline
Cell                                                     & JJs & BC (mA)   &  PG                                                        & RB                                                         & DP     & SS      & TC ($\times 2$)                                           & Other  \\ \hline
Splitter                                                 &   3 & 0.30      &  $N-1$                                                     & $N$                                                        &        & 4       & $2b-2$                                                    & 3      \\
Merger                                                   &   7 & 0.88      &  $N-1$                                                     & 1                                                          &        & 2       &                                                           & 1      \\
DFF                                                      &   6 & 0.72      &                                                            & $N$                                                        &        &         &                                                           &        \\
RTFF                                                     &  13 & 0.808     &                                                            &                                                            &        & 1       & $2b$                                                      &        \\
T1                                                       &  12 & 0.74      &                                                            &                                                            & 1      &         &                                                           &        \\
NDRO                                                     &  33 & 3.464     &                                                            &                                                            &        & 2       &                                                           &        \\
AND                                                      &  14 & 1.428     &                                                            &                                                            & 1      &         &                                                           &        \\ \hline \hline
\begin{tabular}[c]{@{}c@{}}Total BC\\ (mA) \end{tabular} &     &           & \begin{tabular}[l]{@{}r@{}}$1.18N$\\ $-1.18$ \end{tabular} & \begin{tabular}[l]{@{}r@{}}$1.02N$\\ $+0.88$ \end{tabular} & $2.17$ & $10.7$  & \begin{tabular}[r]{@{}l@{}}$2.22b$\\ $-0.60$ \end{tabular} & $1.78$ \\
Serial input                                             &     &           &  No                                                        & Yes                                                        & Yes    & No      & No                                                        & Yes    \\ \hline
\end{tabular}
\end{table}

\begin{figure*}[tb]
  \centering
  \includegraphics{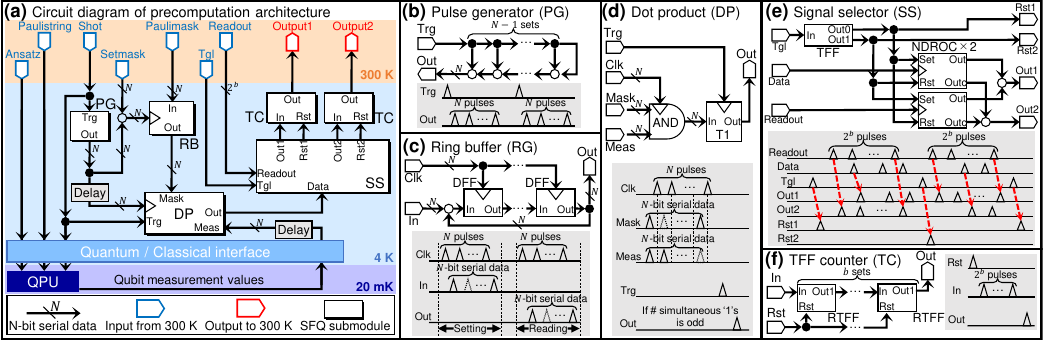}
  \caption{Circuit diagram of our precomputation SFQ circuit}
  \label{fig:circuit_diagram}
\end{figure*}

We first designed our counter-based precomputation architecture for VQE with GM using the existing, well-developed rapid SFQ (RSFQ) cell library~\cite{detail_of_cell_library_ADP2}.
Then we explain that the proposed architecture can realize precomputation without GM.

The library is with a conventional logic family of RSFQ targeting a niobium nine-layer, 1.0-$\mu$m fabrication technology for 4-K temperature operation.
Table~\ref{tab:circuit_configuration} summarizes the SFQ logic gates used in this work with their JJs and bias current (BC) values.

Figure~\ref{fig:circuit_diagram}~(a) shows the circuit diagram of our architecture for the case with GM.
Note that the figure illustrates the circuit for one Pauli string, and our entire architecture consists of the $K$ circuits.
We suppose our circuit for each Pauli string receives $N$-bit `Paulimask' input.

As shown in Fig.~\ref{fig:circuit_diagram}~(a), our architecture consists of five submodules, and Figs.~\ref{fig:circuit_diagram}~(b) to (f) provide the standard cell-level designs of the submodules with their timing chart (gray area).
Table~\ref{tab:circuit_configuration} summarizes the standard cells used in this paper and the detailed configuration of the circuit.

\noindent
\textbf{Pulse generator (PG)}: 
The PG submodule consists of $N-1$ mergers and splitters, and it generates $N$ pulses at the `Out' port upon receiving the `Trg' signal.

\noindent
\textbf{Ring buffer (RB)}: 
The RB submodule constitutes a buffer formed by connecting $N$ DFFs in a ring configuration, designed to store an $N$-bit mask corresponding to a Pauli string. 
To set a mask value, $N$-bit serial data representing the mask is inputted to the `In' port with synchronized $N$ clock pulses. 
Note that when setting the mask value, we adjust the delay of the clock and `In' signals so that the first clock pulse arrives before the first pulse of `In', resulting in the first clock pulse being effectively ignored in the circuit operation (See the timing chart of Fig.~\ref{fig:circuit_diagram}~(c)).
Once the mask value is set, inputting $N$ clock pulses triggers the `Out' port to output the mask value as $N$-bit serial data. 

\noindent
\textbf{Dot product (DP)}: 
The DP submodule computes the dot product between an $N$-bit mask corresponding to a Pauli string and $N$-qubit measurement values. 
Those values are inputted as $N$-bit serial pulse sequences synchronized with $N$ clock pulses to drive the AND gate. 
The T1 gate calculates the parity of simultaneous `1's in the $N$-series inputs. 
If it stores an odd parity, it outputs one pulse to `Out' after the `Trg' signal is input. 

\noindent
\textbf{Signal selector (SS)}
The SS submodule distributes inputs from `In' and `Readout' ports to two distinct outputs, `Out1' and `Out2', ensuring proper allocation. 
Specifically, one output channel is dedicated to transmitting the signal received from the `In' input, while the other channel exclusively handles the signal from the `Readout' input. 
Upon receiving the `Tgl' signal, the module switches the ports for the `In' and `Readout' outputs, effectively reversing their roles. Simultaneously, this transition triggers a signal to the appropriate reset output `Rst1' or `Rst2'.
The SS plays a pivotal role in alternately utilizing the two connected TFF counters.

\noindent
\textbf{T-FlipFlop counter (TC)}: 
The TC submodule is a counter composed of $b$ TFFs connected in series.
Each pulse input to `In' increments the counter value by one, and upon receiving $2^b$ pulses, the counter overflows, and MSB is sent as a pulse to the `Out' port.
Note that there are two TCs combined with an SS submodule in our architecture to hide non-MSB sending latency.

\paragraph{Circuit of precomputation without GM}
Note that our circuit for the case without GM explained in Sec.~\ref{subsec:precomputation_circuit} is a part of Fig.~\ref{fig:circuit_diagram}~(a); we apply it to the case without GM by eliminating the SS and TCs and sending the output of `DP' to the room-temperature environment.

\section{Evaluation\label{evaluation}}

\subsection{Power consumption of precomputation}

To implement our architecture with low-power consumption, we use the energy-efficient RSFQ (ERSFQ)~\cite{mukhanov2011energy} where the static power consumption of RSFQ circuit is eliminated. 
In general, power consumption of ERSFQ circuit is estimated based on RSFQ circuit design using the ERSFQ power model~\cite{mukhanov2011energy} as follows:
\begin{align}
P_{\text{ERSFQ}} =2\times \Phi_{0} \times (\text{BC of RSFQ design}) \times (\text{frequency}). \label{eq:ersfq_model}
\end{align}

First, we formulate the execution time of the entire VQE circuit $t_{QC}$ based on the hardware efficient ansatz~\cite{kandala2017hardware} to define the circuit frequency $f$ as $f = 1/t_{QC}$.
We generally denote the time required to apply a quantum gate $ G $ as $t_{G}$. 
$\ResetT$ and $\MeasureT$ represent the reset and measurement operation time per qubit, respectively.
We assume the maximum gate level parallelism of one- and two-qubit gates; $N$ one-qubit gates can be applied in parallel to each qubit, and $N/2$ two-qubit gates can be simultaneously applied to each pair of qubits on an $N$-qubit machine.
Then, we model the $t_{QC}$ as 
\begin{align}
t_{QC} &=
    \ResetT 
    +
    (D+1)(\ryT + \rzT)
    + 
    2\czT 
    +
    \MeasureT. 
  \label{eq:exec-time-bare}
\end{align}
Considering the worst case for the bandwidth with $D = 1$ and using parameters described in references \cite{jokar2022digiq,opremcak2021measurement} as $\ResetT = 100$~ns, $\ryT = \rzT = 10$~ns, $\czT = 60$~ns, and $\MeasureT = 380$~ns, respectively, $t_{QC}$ in Eq.~(\ref{eq:exec-time-bare}) is estimated as $t_{QC} \simeq 100 + 2\times (10+10) + 60 \times 2 + 380 \times 1 = 640$~ns.

Based on the $t_{QC}$ value and RSFQ design in Sec.~\ref{proposal}, we estimate the power consumption of our system when ERSFQ is applied.
Based on the value of $t_{QC}$ and the circuit design in Sec.~\ref{proposal:counter}, we estimate the power consumption of our circuits without and with GM, denoted as $P_{\text{noGM}}$ and $P_{\text{GM}}$, respectively.

As explained in the last paragraph of Sec.~\ref{proposal:counter}, the circuit for the case without GM is obtained by removing the SS and TCs from the circuit shown in Fig.~\ref{fig:circuit_diagram}~(a).
We use flux quantum $\Phi_0$ of $2.068$~fWb and assume that the operating frequency of the entire circuit $f = 1/t_{QC}$ is 1.56~MHz.
Note that the operating frequency of submodules with serial input is $N$ times that of the entire circuit.
The power of the precompute architecture without GM with ERSFQ can be estimated as follows, as a function of the number of qubits $N$:
\begin{align}
  P_{\text{noGM}}(N) 
  &= 2\Phi_{0}f\times (N(1.02N + 4.83) + (1.18N - 1.18)) \notag \\ 
  &\approx \left(6.6N^2 + 39N - 7.6\right)_{[\text{pW}]}.
  \label{eq:nogm_power}
\end{align}

Similar to $P_{\text{noGM}}$, we estimate $P_{\text{GM}}$ as
\begin{align}
  P_{\text{GM}}(N) 
  \approx \left(6.6N^2 + 39N + (14b + 57)\right)_{[\text{pW}]}.
  \label{eq:gm_power}
\end{align}

\subsection{Scalability to the number of qubits\label{evaluation:scalability}}
\begin{figure*}[tb]
  \centering
  \includegraphics{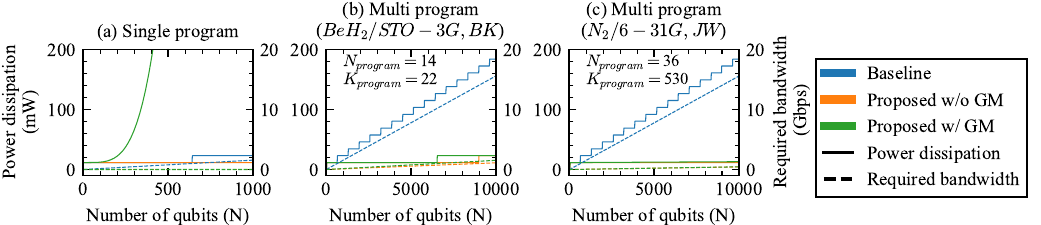}
  \caption{Required bandwidth and power dissipation of baseline and proposed systems without and with GM for (a) single- and (b)---(c) multi-program cases. Power dissipation includes heat inflow and peripherals power consumption. }
  \label{fig:eval_single_multi}
\end{figure*}

Next, we evaluate the trade-off between heat inflow through cables and power consumption with our system in a cryogenic environment. 
We suppose that the systems have stainless steel coaxial cables (UT085 SS-SS) between the 4-K and upper stages, and its heat inflow is 1.0~mW as in Tab.~2 of Ref.~\cite{krinner2019cryosetup}. 
In addition to the heat inflow, we consider the power consumption of the cable peripherals.
We assume the amplifier, which is the main source of peripheral power consumption, is LNF-LNC4\_8C as in Ref.~\cite{krinner2019cryosetup}, and its power consumption is 10.5~mW per cable according to its datasheet. 
The bandwidth per cable is assumed to be 1~Gbps, and the system has a sufficient number of cables to exceed its required bandwidth.
Table~\ref{tab:cable_counter_configuration} summarizes the configuration of cables and our SFQ architecture, 

\begin{table}[bt]
\centering
\scriptsize
\caption{Configuration of cables and SFQ architecture}\label{tab:cable_counter_configuration}
\begin{tabular}{|ll||l|l|}
\hline
\multicolumn{2}{|l|}{}                                                                                                                                & Power dissipation per unit                                                         & Configuration                                                            \\ \hline
\multicolumn{2}{|l|}{\begin{tabular}[c]{@{}l@{}}Coaxial\\ cable\end{tabular}}                                                                         & \begin{tabular}[c]{@{}l@{}}Heat inflow: 1.0 mW\cite{krinner2019cryosetup} \\ Peripherals: 10.5 mW\cite{krinner2019cryosetup}\end{tabular} & One cable per 1 Gbps                                                     \\ \hline
\multicolumn{1}{|l|}{\multirow{2}{*}{\begin{tabular}[c]{@{}l@{}}SFQ\\ arch.\end{tabular}}} & \begin{tabular}[c]{@{}l@{}}Single\\ program\end{tabular} & \begin{tabular}[c]{@{}l@{}}w/o GM: $P_{\text{noGM}}(N)$\\ w/   GM: $P_{\text{GM}}(N)$\end{tabular}               & \begin{tabular}[c]{@{}l@{}}\# units = $1$\\ \# units = $K$               \end{tabular}      \\ \cline{2-4} 
\multicolumn{1}{|l|}{}                                                                     & \begin{tabular}[c]{@{}l@{}}Multi\\ program\end{tabular}  & \begin{tabular}[c]{@{}l@{}}w/o GM: $P_{\text{noGM}}(N_{\text{program}})$\\ w/   GM: $P_{\text{GM}}(N_{\text{program}})$\end{tabular}             & \begin{tabular}[c]{@{}l@{}}\# units = $L$\\ \# units = $LK_{\text{program}}$\end{tabular} \\ \hline
\end{tabular}
\end{table}

First, we show the scalability results for a single program execution case, where one VQE program with $N$-qubit size is executed on an $N$-qubit machine.
Here, we assume $K = N^2$ in the case of performing GM.
Figure~\ref{fig:eval_single_multi}~(a) shows the bandwidth (dashed lines) and power dissipation (solid lines) for the baseline and proposed systems without and with GM during a single-program execution.
The power dissipation of the baseline system (blue solid line) increases in a step-like manner at $N=650$, which indicates an increase in the number of required cables due to the increase in the bandwidth.
For the case without GM, our proposed system has constant bandwidth, which prevents an increase in the number of required cables even as $N$ scales. 
While the power consumption of the proposed circuit scales with $\order{N^2}$ with respect to the number of qubits $N$ as formulated in Eq.~(\ref{eq:nogm_power}), its impact on the power dissipation of the system is minor.

In the case of GM as well, our system can keep the required bandwidth constant by setting counter bit width $b$ to $log\lceil K \rceil$ (green dashed line).
However, the power consumption of our circuit becomes significantly large as $N$ scales (green solid line), due to its $\order{N^4}$ scaling; both $P_{GM}$ and the number of circuits scale with the order of $N^2$.
As a result, in a single-program situation, there is no benefit to implementing the proposed system with GM.

Next, we show the scalability results for a multi-program execution case.
In this situation, we consider the simultaneous execution of a certain VQE program using $N_{\text{program}} (\leq N)$ qubits on an $N$-qubit machine.
Let $L = \lfloor N/N_{\text{program}} \rfloor$ represent the number of VQE programs executed simultaneously, and denote the group size with GM for a given VQE program as $K_{\text{program}}$.
We suppose the ``$BeH_2/STO-3G$'' (``$N_2/6-31G$'') problem with Bravyi–Kitaev (Jordan-Wigner) transformed Hamiltonian in TABLE II of Ref.~\cite{Verteletskyi:2020me}, which is the minimum (maximum) size one shown in the reference.
Note that our architecture is capable of executing a variety of VQE programs simultaneously while we are considering a single type of program here. 

Figure~\ref{fig:eval_single_multi}~(b)---(c) show the bandwidths and power dissipation for the baseline and proposed systems without and with GM in multi-program situations.
While the baseline system shows the same tendency in the single program case, the proposed system maintains a lower bandwidth and power dissipation for both cases with and without GM, even as $N$ scales.
Specifically, in the case of $N=10000$ in Fig.~\ref{fig:eval_single_multi}~(c), where $L = 277$ VQE programs of $N_2/6-31G$ are executed simultaneously, our system reduces the required bandwidth and power dissipation by 97\% and 93\%, respectively, compared to the baseline for both cases without and with GM.

\section{Conclusion\label{sec:conclusion}}
In this paper, we modeled the inter-temperature communication in cryogenic VQE machines and proposed a bandwidth-efficient precomputation architecture using SFQ circuits.
The evaluation result shows that our system successfully reduced the required bandwidth; as a result of this drastic reduction, total power dissipation is significantly reduced in multi-program scenarios on a large-scale cryogenic NISQ machine.

\begin{acks}
This work was partly supported by JST PRESTO Grant Number JPMJPR2015, JST Moonshot R\&D Grant Number JPMJMS2067, JSPS KAKENHI Grant Numbers JP22H05000, JP22K17868, RIKEN Special Postdoctoral Researcher Program.
\end{acks}

\bibliographystyle{ACM-Reference-Format}
\bibliography{ref_arxiv}


\end{document}